\documentclass[debug,overfull,dvips]{epl}

\title{Hydration process of multi-stacked phospholipid bilayers to form giant vesicles}
\shorttitle{Hydration process of phospholipid bilayers to form GV}
\author{M. Hishida\inst{1} \and H. Seto\inst{1} \and N. L. Yamada\inst{2} \and K. Yoshikawa\inst{1}}
\shortauthor{M. Hishida \etal}
\institute{
  \inst{1} Department of Physics, Kyoto University, Kyoto 606-8502, Japan\\
  \inst{2} High Energy Accelerator Research Organization, Tsukuba 305-0801, Japan}

\begin{document}

\maketitle

\begin{abstract}
The hydration process of multi-stacked phospholipid bilayers to form giant vesicles was investigated by time-resolved small angle X-ray scattering. The hydration of lipids in the liquid-crystalline phase was found to proceed in two stages. In the early stage, until about 100 s after hydration, lipid bilayers on a solid substrate swell about 12 $\rm{\AA}$ and reach a quasi-stable state. In the late stage, after several hundreds of seconds, lipid bilayers gradually peel off from the stack. On the other hand, in the case of a lipid in the gel phase, only the early stage is observed. These behaviors correspond to the capacity for the giant vesicle formation depending on the lipid phases. The kinetics of the peeling-off process is discussed in terms of Kramers' formulas.
\end{abstract}

\section{Introduction}

Biomembrane is mainly comprised of phospholipid molecules. Due to the amphiphilic property of phospholipid, these molecules assemble into bilayers in water with their hydrophobic tails facing inward, and form vesicles called liposomes \cite{1}. To investigate a model cell system, it is important to establish a method for creating giant vesicles (GV) with a diameter above 1 $\rm{\mu}$m, which are comparable in size to living cells \cite{2}. Phospholipid molecules usually form small unilamellar vesicles (SUV) or multi-lamellar vesicles (MLV), as described in the literature \cite{6}, through the use of sonication and/or mechanical agitation. On the other hand, a method for forming GVs has not been fully established, and especially the fundamental mechanism that is involved in the formation of GVs is not yet understood \cite{4,5,6}. 

A natural swelling method, in which GVs are spontaneously formed by hydrating multi-stacked dry phospholipid layers on a solid substrate, is quite useful for creating a model cell because no physical or chemical stresses are applied to GVs \cite{4,5,29}. In this method, each lipid bilayer should peel off from the stack of bilayers on a substrate and form a vesicle. Therefore, it is expected that the mechanism of this natural swelling is closely associated with the unbinding transition \cite{8,50,51}. It is to be noted that this mechanism should be discussed as a kinetic unbinding process that occurs through hydration, although the unbinding transition is commonly discussed in the framework of a temperature-induced transition \cite{9,10}. 

Recently, the present authors verified the conditions needed to effectively form giant vesicles from dry phospholipid films through natural swelling \cite{11,34}. The terrace-like morphology of dry phospholipid film and the adequate thermal fluctuation of bilayers in the liquid crystalline phase are necessary for obtaining GVs. On the other hand, a rough morphology is seen for dry lipid films when the lipid is in the gel phase, and no GVs are formed under this condition. These results indicated that the terrace-like morphology of dry lipid film and the liquid-crystalline phase are essential for the formation of GVs. Although the initial and final structures in the vesicle formation are clarified, the hydration process that leads to the formation of GVs from dry films remains to be solved.

Time-resolved small angle X-ray scattering (TR-SAXS) is a powerful tool for the in situ observation of structural changes on the order of nm. Several TR-SAXS studies have been performed on the kinetics of vesicle formation or the unbinding transition. Thimmel et al. investigated the process of swelling from a stack of lipid bilayers. However, they examined this process by increasing temperature (not swelling with water) \cite{13}. Hartung et al. and Pabst et al. have shown that the intensity of Bragg peak due to the regular stacking of bilayers decrease slowly with time when well-aligned phospholipid bilayers on a solid substrate are hydrated by excess water \cite{50,52}. Although their results suggested that the bilayers peel off from the substrate and form vesicles, the process of hydration has not been clarified yet. Further, the kinetics has not been discussed quantitatively \cite{51}.

In the present letter, we report the hydration process of phospholipid dry film in detail as examined by TR-SAXS. The process was found to have two stages in the formation of giant vesicles. Water molecules penetrate between lipid bilayers in the early stage, and each bilayer peels off from the stack of bilayers in the late stage. The kinetics of the late stage is discussed quantitatively in terms of Kramers' formulas.

\section{Materials and Methods}

1,2-Dioleoyl-sn-glycero-3-phosphocholine (DOPC, SIGMA-ALDRICH) and 1,2-dipalmitoyl-sn-glycero-3-phosphocholine (DPPC, Wako Pure Chemical Industries) were obtained in powder form and used without further purification. DOPC is in the liquid-crystalline phase in both dry and wet conditions at room temperature, while DPPC is in the gel phase under the same conditions \cite{33}. These lipids were dissolved in an organic solvent composed of dehydrated chloroform and dehydrated methanol (2:1 v/v) as mother solutions (both from NACALAI TESQUE) with molecular sieves (NACALAI TESQUE). These mother solutions (10 mM) were stored at -30 ${}^\circ\! \rm{C}$ before use. Each mother solution (10 $\rm{\mu}$L) was dropped onto a capillary glass tube that had an outer diameter of 1.5 mm. By evaporating the organic solvent overnight in a vacuum, dry phospholipid film 8 mm long remained on the surface of the glass tube. Under these conditions, about a hundred lipid bilayers were stacked on the glass with the total thickness being 400-500 nm which was estimated by X-ray reflectivity measurement \cite{34}.

\begin{figure}
\onefigure[width=.6\linewidth]{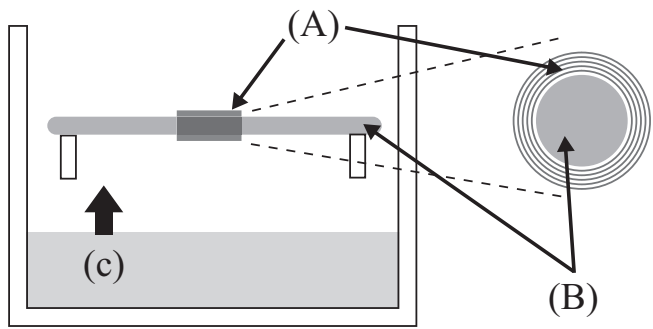}
\caption{Schematic image of the sample cell used to study the hydration of lipids by TR-SAXS. Dry phospholipid film (A) was prepared on the outer surface of a capillary glass tube (B) hung in an acrylic cell. The upper part of the lipid film was radiated by an incident X-ray beam. Pure water (C) was automatically supplied from a water tank (not shown) and the phospholipid film was hydrated after the SAXS measurement was started.
\label{fig1}}
\end{figure}

To observe the hydration process in detail, TR-SAXS experiment was carried out at BL40B2, SPring-8, JASRI (the Japan Synchrotron Radiation Research Institute), Japan. The wavelength of the X-ray was 1 $\rm{\AA}$ and the diameter of X-ray at the sample was 0.4 mm. The scattered X-ray beam was detected by CCD. The sample-to-detector distance was about 1 m, which was calibrated using a standard sample (lead stearate). The observed momentum transfer ranged over $0.018 \leq {\rm{q}} \leq 0.63 \, \mathrm{\AA^{-1}}$. The samples were kept in a special cell, as shown in Fig. \ref{fig1}. The capillary glass tube with dry phospholipid film was hung in an acrylic cell with windows for X-ray made of Kapton film. Pure water (MilliQ) was automatically supplied into the cell and the water hydrated the dry lipid film. TR-SAXS measurements were started just before hydration. All the measurements were performed at room temperature.

\section{Results}

\begin{figure}

\onefigure[width=.9\linewidth]{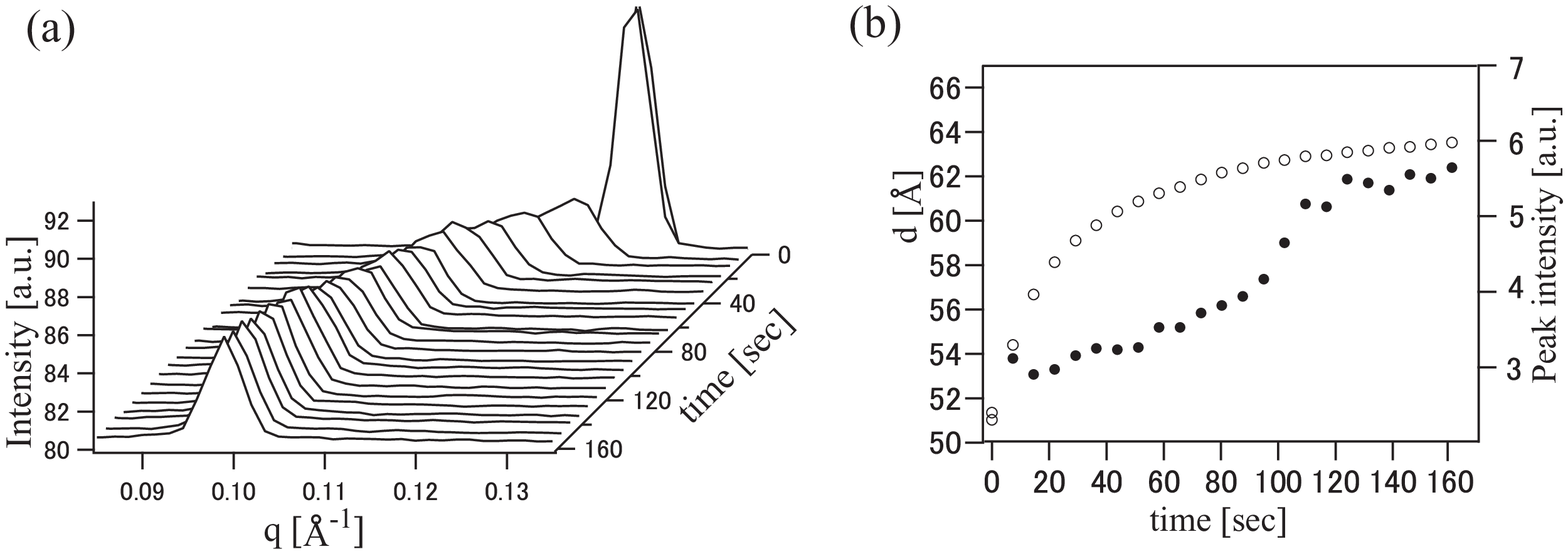}
\caption{(a) Time-dependence of the SAXS profile observed in the early stage of hydration of DOPC dry film. A Bragg peak corresponding to the regular stacking of lipids bilayers shifts to lower-$q$ and decreases in intensity just after hydration, and reaches $q \approx 0.099 \, \mathrm{\AA}^{-1}$. (b) The time-dependence of the lamellar repeat distance $d = 2 \pi / q$ ($\circ$) obtained from the position, and the peak intensity of the Bragg peak ($\bullet$). $d$ is saturated to be $\approx$ 64 $\rm{\AA}$ at around 100 sec. The peak intensity decreases just after hydration and then increases with time.
\label{fig2}}
\end{figure}

\begin{figure}
\onefigure[width=.9\linewidth]{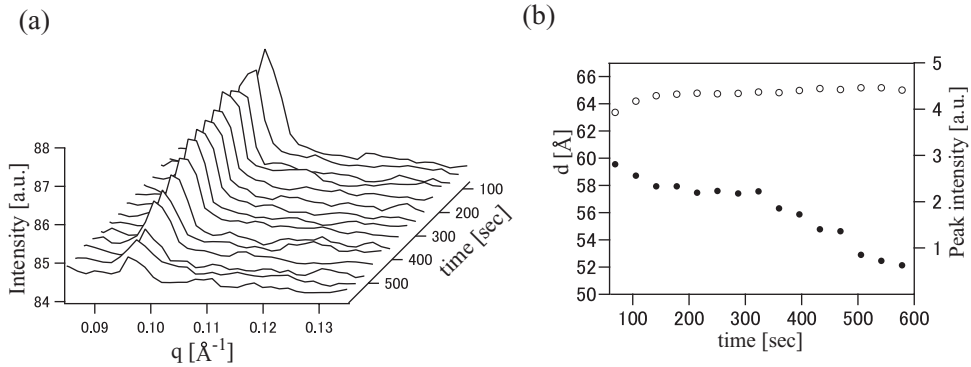}
\caption{(a) The TR-SAXS profile observed in the late stage of hydration of the DOPC dry film. The position of the Bragg peak does not change and only the peak intensity decreases with time. (b) The time-dependence of the lamellar repeat distance $d $ and the peak intensity obtained from the SAXS profile. 
\label{fig3}}

\end{figure}

The time-dependence of the SAXS profiles in the hydration of the DOPC film are shown in Figs. \ref{fig2} and \ref{fig3}. The early stage (0 - 100 s after hydration) and the late stage (100 - 600 s) of hydration are shown separately in Fig. \ref{fig2} and Fig. \ref{fig3}, since the time scales of these processes are different, i.e., the time-step of the measurement in the early stage was 72.7 ms and that in the late stage was 727 ms. 

A large difference in behavior was noted between the early stage and the late stage. A sharp peak corresponding to the regular stacking of lipid bilayers tends to shift to a lower-$q$ in the early stage. Before hydration (0 s), the peak is at $q = 0.12 \, \mathrm{\AA^{-1}}$. Within 0 - 20 s after hydration, the intensity of the Bragg peak from the regular stacking of dry lipid layers decreases and the position of the peak shifts slightly and a broadening occurs. Next, from 20 to 100 s, the peak position moves inverse-exponentially down to $q=0.0989 \pm 0.0001 \, \mathrm{\AA^{-1}}$, which corresponds to the lamellar repeat distance $d=64 \, \mathrm{\AA}$, and the peak intensity increases simultaneously. The value $d=64 \, \mathrm{\AA}$ is 1-2 $\rm{\AA}$ larger than the lamellar repeat distance of DOPC MLV \cite{14}. Although the reason on the difference is not clear yet, it may be attributed to any experimental errors, or an effect of the substrate, or some intrinsic origins as the vapor pressure paradox \cite{53}. Anyway, it is apparent that the final value of $d$ is almost the same as that of MLV. In this stage, the regularity of the bilayer stacking initially breaks, and then recovers at about 100 s after hydration. The result means that, in the early stage, the lipid film on a solid substrate swells and reaches a quasi-equilibrium state, due to the penetration of water molecules among lipid bilayers. The integrated intensity of the peak corresponding to the number of layers in the film does not change with time (data not shown). This indicates that multi-stacked DOPC bilayers do not peel off from the substrate in this process. 

Figure \ref{fig3} shows the time-dependence of the SAXS profile in the late stage, from 100 s to 600 s after hydration began, for the DOPC film. In this process, the lamellar repeat distance $d$ for the lipid bilayers on the substrate remains almost constant and the intensity of the peak decreases monotonically. In addition, the peak width does not change and the integrated intensity decreases as the peak intensity. Therefore, the peak intensity corresponds to the number of stacked bilayers. The time dependence of the number of stacked bilayers is shown in Fig. \ref{fig4} (b) ($\bullet$), which estimated from the number of stacked bilayers in the initial condition (dry film). This behavior suggests that the lipid bilayers are unbound and peel off from the stack in this stage. Based on these results, we can conclude that the hydration process of dry DOPC film proceeds in two stages. In the early stage from 0 s to about 100 s after hydration begins, water molecules penetrate between bilayers and the multi-lamellar structure reaches a quasi-stable state. In the following stage, the lipid bilayers gradually peel off from the multi-lamellar stack (maybe one by one), and form giant vesicles. The behavior of the late stage is consistent with the results by Hartung et al. and Pabst et al. \cite{50,52,51}. However, the present process is much faster than their cases. It should be noticed that the early stage is observed for the first time.

To compare the hydration process with that in the gel phase, DPPC dry film was investigated by the same method. In this case, only the early stage, a Bragg peak corresponding to the regular stacking of bilayers shifts to lower-$q$, is observed. The final repeat distance $d$ is 66 \AA, which is also 1-2 $\rm{\AA}$ larger than the lamellar repeat distance of MLV \cite{14}. The peak intensity does not change until 350 sec after hydration begins. This means that the number of stacked bilayers does not change and no bilayers peel off from the stack on the substrate. This result corresponds to the fact that no GVs is formed by the hydration of dry DPPC film in the gel phase at room temperature \cite{11,34}.

\section{Discussion}

\begin{figure}
\onefigure[width=.9\linewidth]{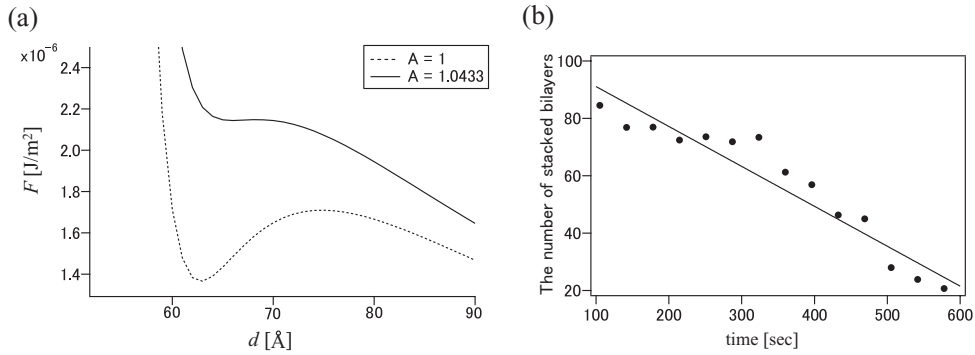}
\caption{(a)Calculated free energies of regular stacking of lipid bilayers. The dashed line indicates the free energy of an inner layer in the stack ($A = 1$) and the solid line is the free energy of the outermost layer with effective Helfrich repulsion. The energy barrier becomes small enough to exhibit spontaneous unbinding by thermal fluctuation when the factor of the effective Helfrich repulsion is about 4 \% larger than the normal value. (b) Experimentally estimated decay rate of the number of stacked bilayers ($\bullet$) and the calculated rate with the Kramers' method (solid line) (See the discussion.).
\label{fig4}}

\end{figure}

The equilibrium distance $d$ of lipid bilayers has been interpreted in terms of the balance of interactions between bilayers. Since the present systems do not include salts, van der Waals attractive interaction and the repulsive force due to the binding of water molecules at hydrophilic headgroups of lipid molecules, and steric repulsion that originates from thermal fluctuations of lipid bilayers should be taken into consideration \cite{15,16}. Therefore, the free energy per unit area of bilayers depending on $d$ can be written as,
$$
F(d) = F_{vdW} (d) + F_{hyd} (d) + F_{Hel} (d)
$$
$$
F_{vdW} (d) = - \frac{H_1}{12 \pi} \left[\frac{1}{(d-a_\bot)^2} - \frac{2}{d^2} + \frac{1}{(d+a_\bot)^2} \right]
$$
$$
F_{hyd} (d) = P_{hyd} \lambda \mathrm{exp} \left(- \frac{d-a_\bot}{\lambda} \right)
$$
$$
F_{Hel} (d) = 0.42 \frac{(k_B T)^2}{ K_c (d-a_\bot)^2}
$$
where $H_1$ is the Hamaker constant, $a_\bot$ is the thickness of a bilayer, $P_{hyd}$ is a prefactor of $F_{hyd}$, $\lambda$ is the decay length of a hydration layer, $K_c$ is the bending rigidity of a bilayer, and $k_B$ is the Boltzmann constant \cite{16}. This free energy profile is calculated for a layer in regularly multi-stacked bilayers, and is also applicable to the bilayers in MLV \cite{15,16}. The calculated free energy profile of DOPC multi-layers as a function of $d$ is shown in Fig. \ref{fig4} (a) (Dashed line. The values of the coefficients are taken from \cite{17}). The observed early stage is interpreted as a relaxation process to the local minimum at $\approx 64 \, {\rm{\AA}}$ and the late stage is a diffusion (unbinding) process from the local minimum. In this case, the energy barrier $\Delta F$ (energy difference between the local minimum and maximum) is $\sim 10^3 k_B T$, and the unbinding of bilayers due to thermal fluctuation, i.e. peeling off of bilayers, should be too slow to observe. Thus, each phospholipid bilayer in multi-stacked bilayers is bound tightly as in a stable state. It should be noted that the present authors previously showed that a collaboration with other repulsive forces is necessary for unbinding \cite{11,40}. 

The case of the outermost layer of a stack of lipid bilayers could be different from that in the inner layers, since one side of the outermost bilayers is not restricted by other bilayers, the edge of the layer is exposed to water, and water can flow into bilayers. Thus, here we assume that the effective Helfrich repulsion of the outermost bilayer is slightly larger than the theoretical value:
$$
F(d) = F_{vdW} (d) + F_{hyd} (d) + A \, F_{Hel} (d)
$$
where $A$ accounts for the effective repulsion of the outermost bilayer ($A > 1$). Such an assumption allows us to explain the experimental evidence of the spontaneous unbinding to form vesicles in the late stage. The one-dimensional Smoluchowski equation is 
$$
\frac{\partial f}{\partial t} = \frac{1}{\gamma m} \frac{\partial}{\partial d} \left( -Kf + k_B T \frac{\partial f}{\partial d} \right)
$$
where $K = - {\partial F}/{\partial d}$. $\gamma$ and $m$ are a resistance coefficient and the mass of a lipid bilayer, respectively. The outermost bilayer is assumed to be disc-shaped with diameter $R$, and the resistance coefficient $\gamma$ is derived as $\gamma = 16\mu R/m $ using Stokes' law, where $\mu$ is the viscosity of water \cite{19}. 

In the present explanation, it is reasonable to assume that the diffusion behavior of the outermost bilayer remains almost constant because the number of stacked bilayers is sufficient within the range of this experiment, i.e., we can consider the next outermost bilayer after a neighboring bilayer has left to form a GV. This assumption is consistent with the experimental evidence that the peak width, which corresponds to the distribution of the mean repeat distance of the stacked bilayers, remained constant in the late stage and the peak height decreased monotonically at an almost constant rate. Thus, we can calculate the Smoluchowski equation analytically by Kramers' method \cite{18} and the number of bilayers passing through the energy barrier per unit time is
$$
r \approx \frac{1}{\sqrt{2 \pi}} \frac{c c' m}{16 \mu R} \mathrm{exp} \left( - \frac{\Delta F}{k_B T} \right)
$$
where $c$ and $c'$ are curvature radii at the minimum and maximum of the free energy, respectively. We can obtain $c$, $c'$, $\Delta F$ from the calculated free energy. With these coefficients, by substituting $R = 1 \, [\rm{\mu m}]$, $m = 2.8 \times 10^{-15} \, [\rm{g}]$, $\mu = 8 \times 10^{-4} \, [\rm{Pa \cdot s}]$, and $k_B T = 4.14 \times 10^{-21} \, [\rm{J}]$, we can estimate $r$ numerically. In the case of $A = 1.0433$, the time-dependence of the number of stacked bilayers calculated with this assumption is consistent with the experimental result, which is depicted in Fig. \ref{fig4} (b) by the solid line. The free energy profile in this case is shown as a solid line in Fig. \ref{fig4} (a). From this estimation, it is concluded that a few-percent increase in the Helfrich repulsion only for the outermost bilayer triggers the unbinding of a bilayer from a stack of bilayers. 

The same discussion can be applied in the case of the DPPC. Since it is in the gel phase at room temperature, the Helfrich repulsion is much smaller than in the case of DOPC; $\Delta F$ is $\sim 10^4 k_B T$ even if the effective Helfrich repulsion of the outermost bilayer is assumed to be two times larger than the normal Helfrich repulsion ($A = 2$). This means that bilayers cannot be spontaneously unbound by thermal fluctuation. 

\section{Conclusion}

We summarize the process of natural swelling from a stack of lipid bilayers on a solid substrate as follows. In the early stage of the hydration, water molecules penetrate within the stack of lipid bilayers and the multi-stacked bilayers reach a quasi-equilibrium state. Following this process, bilayers gradually peel off and form vesicles. In this late stage, only the outermost bilayer is unbound due to thermal fluctuation and peels off from the stack. This process can occur only when the lipid is in the liquid-crystalline phase with a terrace-like morphology on a solid substrate. In the gel phase, lipid bilayers remain stacked and GVs are not formed.

\acknowledgements
The authors are indebted to Dr. K. Inoue and Dr. S. Sasaki at JASRI and Dr. M. Takenaka and Dr. H. Kitahata at Kyoto Univ. for their supports with the experiment and the data analysis. The SAXS experiments were performed under the approval of JASRI (No. 2004B0520-NL2b-np). This work was supported by a Grant-in-Aid for Scientific Research (No. 17540382), by a Grant-in-Aid for the 21st Century COE ``Center for Diversity and Universality in Physics'' from the Ministry of Education, Culture, Sports, Science and Technology (MEXT) of Japan, and by the Yamada Science Foundation.

\end{document}